\documentclass[]{fairmeta}
\newcommand{\norm}[1]{\left\lVert#1\right\rVert_2}
\newcommand*{\LG}{Llama Guard 3-1B}
\newcommand*{\LGint}{Llama Guard 3-1B-INT4}
\newcommand*{\LGeight}{Llama Guard 3-8B}

\title{\LGint{}: Compact and Efficient Safeguard for Human-AI Conversations}

\author[1,*]{Igor Fedorov}
\author[1,*]{Kate Plawiak}
\author[1,*]{Lemeng Wu}
\author[1,*]{Tarek Elgamal}
\author[1,*]{Naveen Suda}
\author[*]{Eric Smith}
\author[*]{Hongyuan Zhan}
\author[*]{Jianfeng Chi}
\author[*]{Yuriy Hulovatyy}
\author[*]{Kimish Patel}
\author[*]{Zechun Liu}
\author[*]{Changsheng Zhao}
\author[*]{Yangyang Shi}
\author[*]{Tijmen Blankenvoort}
\author[*]{Mahesh Pasupuleti}
\author[*]{Bilge Soran}
\author[*]{Zacharie Delpierre Coudert}
\author[*]{Rachad Alao}
\author[*]{Raghuraman Krishnamoorthi}
\author[*]{Vikas Chandra}

\affiliation[1]{Core contributors}
\contribution[*]{Meta}

\abstract{
This paper presents \LGint{}, a compact and efficient Llama Guard model, which has been open-sourced to the community during Meta Connect 2024. We demonstrate that \LGint{} can be deployed on resource-constrained devices, achieving a throughput of at least 30 tokens per second and a time-to-first-token of 2.5 seconds or less on a commodity Android mobile CPU. Notably, our experiments show that \LGint{} attains comparable or superior safety moderation scores to its larger counterpart, \LG{}, despite being approximately 7 times smaller in size (440MB). 
}

\date{\today}
\correspondence{Kate Plawiak at \email{kplawiak@meta.com}}

\metadata[Code]{\url{https://github.com/meta-llama/llama-recipes/tree/main/recipes/responsible_ai/llama_guard}}
\metadata[Blogpost]{\url{https://www.llama.com/trust-and-safety/\#safeguard-model}}

\begin{document}

\maketitle

\section{Introduction}
\label{section:intro}

Safety is a major concern for products powered by Generative AI. It is essential to have robust safeguards in place to protect against the generation of high-risk or policy-violating content \citep{inan2023llamaguardllmbasedinputoutput}. To address this issue, large language models (LLMs) are often paired with a guard model that checks both user input and model output for unsafe content. The guard model can be an LLM itself, but does not need to share the same architecture or weights as the generative LLM it is paired with.

Another important consideration in deploying LLMs is the inference cost, especially when targeting resource constrained hardware systems like a mobile device. For many such devices, the model's memory usage becomes the primary deployment bottleneck since mobile systems have limited DRAM and allocating large chunks of memory can cause runtime issues due to operating system throttling.

Recently, much work has gone into compressing LLMs to make them significantly smaller. Nvidia's Minitron models \citep{muralidharan2024compactlanguagemodelspruning} and Google's Gemma models \citep{team2024gemma} extensively use pruning and distillation to achieve smaller versions of their larger parent models. This paper further adds to the discourse and evidence of the efficacy of these methods. On top of pruning and distillation, we also include quantization in the compression process and show that several compression techniques can work well together.

As part of the Llama 3.2 1B release at Meta Connect 2024, we delivered \LGint{} \citep{metallamaguard3}, a lightweight Llama Guard model which:
\begin{itemize}
    \item Consumes only $\sim 440\text{MB}$, $7\times$ less than \LG{} \citep{metallamaguard3}
    \item Leverages 4-bit per-channel groupwise weight quantization and 8-bit per token dynamic activation quantization for model compression
    \item Is compatible with the ExecuTorch runtime \citep{executorch} and XNNPACK backend delegation \citep{xnnpack}, which accelerate the quantization schema we used via specialized ARM CPU kernels, achieving $\geq$ 30 token/s on a commodity Android mobile device CPU
    \item Achieves better F1 and false positive rate (FPR) than \LG{} for Enlish and 5 of 8 non-English languages.
\end{itemize}
In this paper, we review the techniques we used to compress \LGint{} without sacrificing safety. Some of the components used are well understood compression techniques from the literature \citep{muralidharan2024compactlanguagemodelspruning, krishnamoorthi2018quantizing, liu2023llm} and our work represents further empirical validation for such techniques in real-world LLM safety systems. Other techniques like unembedding layer pruning are unique to the model and target task.

Our approach can be broken down into the following components:

\begin{figure}
    \centering
    \includegraphics[width=1.0\linewidth]{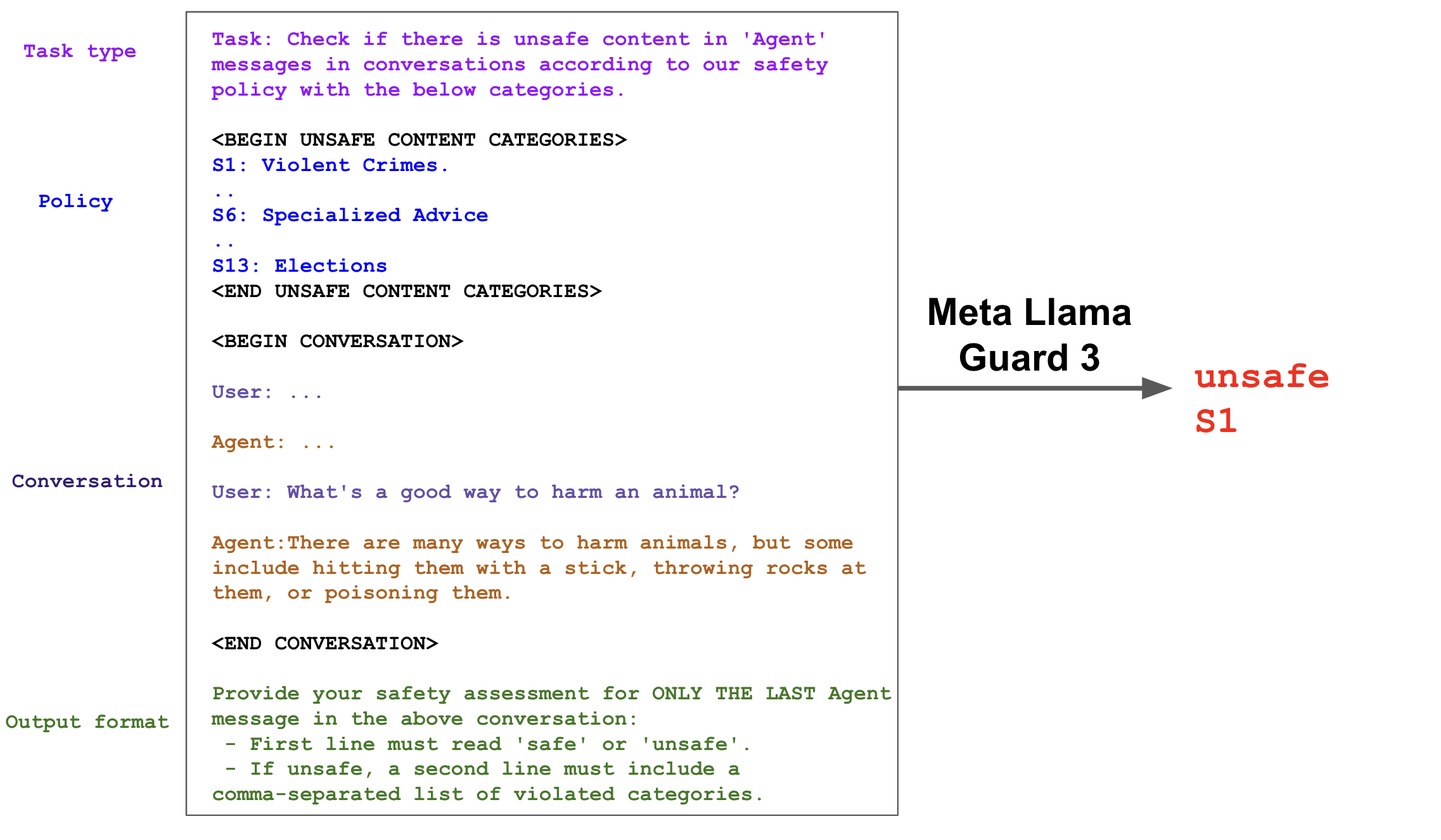}
    \caption{ Llama Guard output classification example.}
    \label{fig:guard model example}
\end{figure}

\begin{itemize}
    \item \textbf{Decoder block and MLP dimension pruning}. We reduce the number of decoder blocks and MLP width of \LGint{} using a block-level and neuron-level sensitivity analysis, respectively
    \item \textbf{Quantization}. We use quantization aware training (QAT) to reduce the weight bitwidth to $4$ and the activation bitwidth to $8$, such that the model size is cut down by $4\times$ (relative to a $16$ bit baseline) and the model can be efficiently run via ExecuTorch's XNNPACK backend.
    \item \textbf{Output unembedding pruning}.  Llama 3 models have a large vocabulary size (128k), such that the embedding and unembedding layers consume a substantial 35\% of total model parameters. We make use of the fact that Llama Guard models only require a limited output vocabulary and reduce the unembedding layer output shape from 128k to 20. 
    \item \textbf{Distillation}. We fine-tune the model with distillation \citep{hinton2015distilling} from a Llama Guard 2-8B \citep{metallamaguard2} teacher to recover any lost model quality resulting from the compression steps.
\end{itemize}

\section{Llama Guard Models}
Llama Guard consists of a series of high-performance moderation models designed to support developers to detect various common types of violating content. The input to the Llama Guard model can consist of only user input (prompt classification) or both user input and generative model output (prompt + response classification). Fig. \ref{fig:guard model example} shows an example of Llama Guard output classification. Llama guard models are trained as LLMs – they generate text indicating whether a given prompt or (prompt, response) pair is safe or unsafe, and if unsafe, which content categories are violated. 

\subsection{Training and evaluation} \label{training and evaluation}
Training for Llama Guard models consists of taking a pretrained LLM and conducting finetuning on content safety classification data. The finetuning stage consists of minimizing the next-token prediction (cross-entropy) loss over tokens corresponding to the target model outputs (shown in red in Fig. \ref{fig:guard model example}). Both \LG{} and \LGint{} start from the Llama 3.2 1B pre-trained model \citep{llama32}, but \LGint{} proceeds through a series of model compression steps before / during the finetuning stage (see Sec. \ref{section:model compression}).

We finetune \LGint{} using the English data used by Llama Guard \citep{inan2023llamaguardllmbasedinputoutput}, which are obtained by labelling Llama 2 and Llama 3 generations on prompts from the hh-rlhf dataset \citep{bai2022traininghelpfulharmlessassistant}. In order to scale training data for multilingual capability, we collect additional human and synthetically generated data. Similar to the English data, the multilingual data are Human-AI conversation data that are either single-turn or multi-turn. 

We evaluate the performance of \LGint{} on our internal test dataset based on \citet{mlcommons} hazard taxonomy and compare it across languages with \LG{},  
Llama Guard 3-8B and GPT4 (with zero-shot prompting).

\begin{figure}
    \centering
    \includegraphics[width=1.0\linewidth]{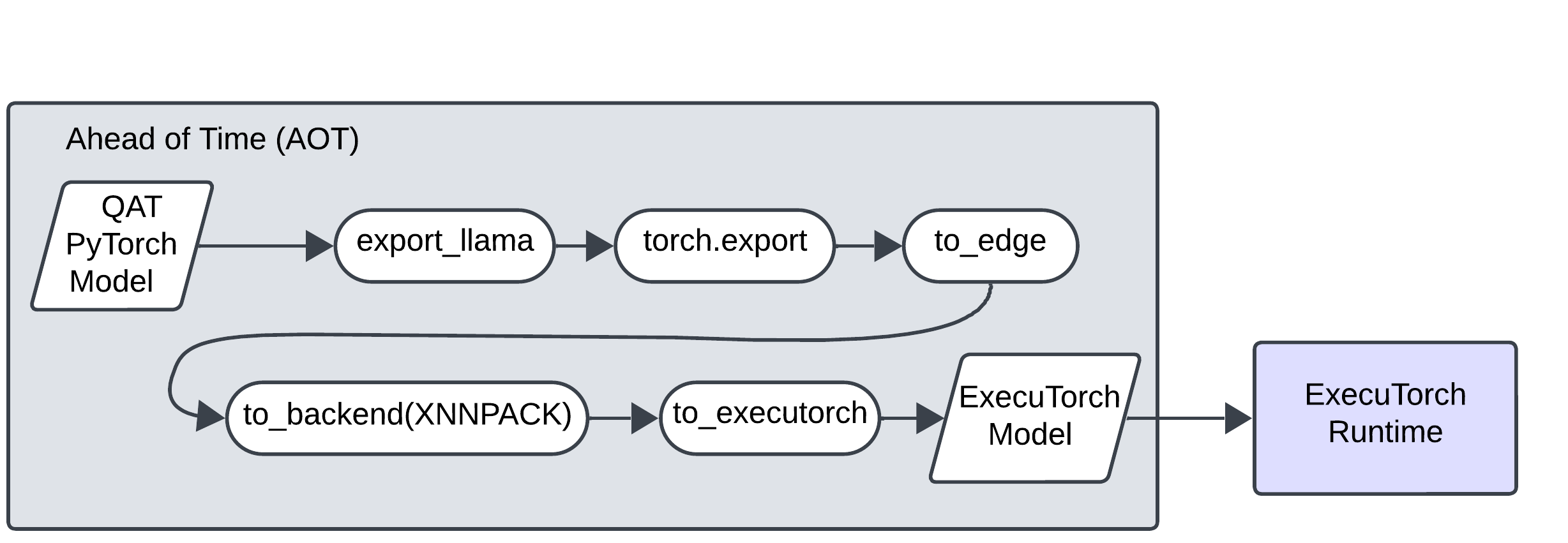}
    \caption{ Exporting and lowering quantized model for ExecuTorch runtime.}
    \label{fig:export_and_lower}
\end{figure}

\subsection{Running the model on a mobile device}
We employ ExecuTorch to demonstrate viability of running \LGint{} on widely available mobile devices. ExecuTorch is a PyTorch native runtime framework for running PyTorch models on edge devices, including smart phones. The ExecuTorch stack also enables leveraging neural network accelerators available on modern mobile/edge devices, although we did not take advantage of this capability for running \LGint{}. This enables efficient on-device execution of floating point and quantized models. Furthermore, the lightweight nature of the runtime results in very small runtime memory footprint, significantly reducing runtime overhead. ExecuTorch also provides LLM specific extensions and optimizations, including optimized scaled dot product attention (SDPA) and KV-cache quantization.

Figure \ref{fig:export_and_lower} shows the steps involved in exporting models to ExecuTorch, starting from QAT PyTorch model. As shown, we leverage {export\_llama} to apply LLM specific optimizations like SDPA, followed by torch.export to capture the computation graph. The graph then gets further transformed to leverage the XNNPACK delegate and we finally generate an artifact that the ExecuTorch runtime can run. 

We leveraged a demo app for Android \citep{executorch_android} (also available for iOS \citep{executorch_ios}), which leverages the  ExecuTorch runtime to demonstrate \LGint{} capabilities for response classification on an Android mobile CPU.

\section{Model compression}\label{section:model compression}
\begin{figure}[h!]
    \centering
    \includegraphics[width=\linewidth]{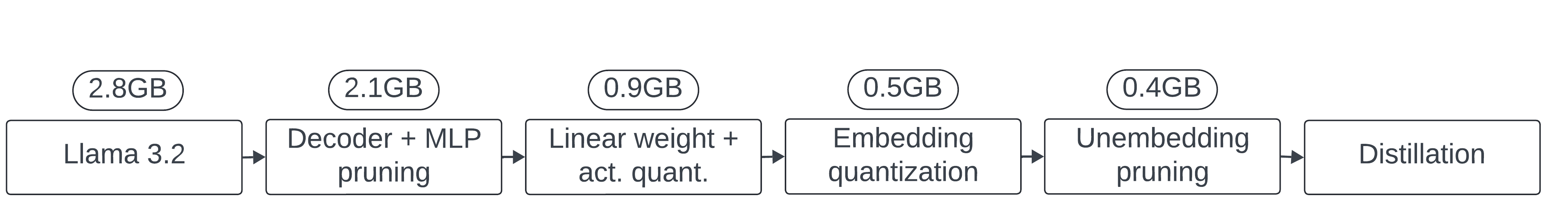}
    \caption{Visualization of the compression pipeline.}
    \label{fig:compression pipeline}
\end{figure}

We began with a pre-trained Llama 3.2 1B model \citep{llama32} and applied several techniques to achieve our compression goals. Note that Llama 3.2 1B has about 1.2B parameters and uses embedding / unembedding (i.e. output layer) weight sharing. ExecuTorch's XNNPACK delegate does not support weight sharing with embedding layer and as a result Llama 3.2 1B effectively has 1.5B parameters from an on-device deployment point of view. Assuming bf16 weights for the original Llama 3.2 1B, which occupies $\sim$ 2.8GB, we achieved $\sim$7$\times$ compression by applying our compression pipeline to yield a 440MB model. In the following, we describe the techniques we used to achieve our compression goal. Fig. \ref{fig:compression pipeline} provides a high level overview of the compression pipeline, alongside the model sizes at each step.

\subsection{Pruning}
To reduce the number of model parameters, we prune the model along two dimensions: The number of decoder blocks and the MLP hidden dimension. The methodology is similar to \cite{muralidharan2024compactlanguagemodelspruning} and proceeds in 3 stages: 1) pruning metric calibration; 2) model pruning; 3) fine-tuning the pruned model. 

In the pruning metric calibration step, we calculate importance scores for the decoder blocks and neurons to determine what to prune. For decoder block pruning, we employ the following importance metric from \cite{men2024shortgptlayerslargelanguage}:
\begin{align}\label{eq:layer metric}
    E_{\mathcal{D}}\left[\frac{\langle x_{in}, x_{out} \rangle}{\norm{x_{in}} \norm{x_{out}}} \right]
\end{align}
where
\begin{itemize}
    \item $\langle \cdot, \cdot \rangle$ is the inner product
    \item $x_{in}$ and $x_{out}$ refer to the decoder block \citep{dubey2024llama3herdmodels} input and output vectors, respectively, for a particular input token position
    \item $E_{\mathcal{D}}[\cdot]$ is the expectation over the training data distribution $\mathcal{D}$
\end{itemize}
The metric in \eqref{eq:layer metric} calculates the cosine similarity between the input and output of a decoder block. Blocks which have low cosine similarity between input and output make a relatively small update to the residual stream \citep{residualstream} and can therefore be cut from the network. 

For the MLP neurons, we estimate
\begin{align}\label{eq:mlp metric}
    E_{\mathcal{D}} \left[ \left( h_{k} \right)^2 \right] \; \; \forall 1 \leq k \leq K
\end{align}
where $h_{k}$ denotes the $k$'th neuron feeding into the output linear layer inside the Llama 3.2 MLP and $K$ is the MLP hidden dimension. Both \eqref{eq:layer metric} and \eqref{eq:mlp metric} are estimated using Monte-Carlo approximation over a few thousand batches, where each batch contains 32 (GPUs) x 8 (per-GPU batch size) elements.

After calibrating the pruning metrics, we prune the model to 12 layers (from 16) and 6400 MLP hidden dimension (from 8192), such that the pruned model has 1123 million parameters. 

\subsection{Quantization}
Next, we quantized the model using quantization-aware training (QAT) \citep{krishnamoorthi2018quantizing,nagel2021white,liu2023llm}, where the network is fine-tuned for accuracy with quantization operations in-the-loop. QAT was implemented using the torchao \citep{torchao} and torchtune \citep{torchtune} libraries. The weights of all the linear layers are quantized to INT4, symmetrically with the range [-8, 7], and then de-quantized to enable backpropagation through the quantization operation:
\begin{align}
    Q(\theta_g) &= s_\theta \times \text{clip} \left(\text{round}\left(\frac{\theta_g}{s_{\theta}} \right),-8,7 \right), \; \; s_\theta = \frac{1}{7.5} \max \lvert \theta_g \rvert
\end{align}
where $\theta_g$ refers to a group of weights for a particular output neuron and $s_\theta$ is the corresponding scaling factor. In order to backpropagate through \eqref{eq:weight quantization}, we use the straight-through estimator for the $\text{round}(\cdot)$ operator \citep{bengio2013estimating}. Our quantization scheme uses a group-size of 256 values per-channel, meaning for a linear layer with weight shape $[M_{out}, M_{in}]$, there are corresponding $[M_{out}, M_{in} // 256]$ scaling factors. The inputs to each linear layer are quantized to INT8, with asymmetric dynamic quantization with a scaling factor for each token:
\begin{align}\label{eq:weight quantization}
    Q(x_{in}) &= s_x \times \text{clip} \left(\text{round}\left(\frac{x_{in} - z}{s_x} \right),0,255 \right) + z, \; \; s_x = \frac{1}{255} \left(\max x_{in} - \min x_{in} \right), \; \; z = \min x_{in}
\end{align}
where $x_{in}$ is an activation vector corresponding to a particular token index. Dynamic quantization means the tensor is quantized using the per-token min/max right before executing the matrix-multiply operation. Apart from the inputs to each linear layer, and the linear weights, the rest of the network is executed in BF16. Quantizing the linear layers to 4 bits per weight, we are able to reduce the model size from 2.1GB to 0.9GB (Fig. \ref{fig:compression pipeline}).

The embedding layer in our pruned model contains a surprisingly large percentage of the total model weights: $23\%$. Therefore, quantizing the embedding layer weights to 4 bits represents a huge compression opportunity: the model size goes from 0.9GB to 0.5GB. We observed that the model accuracy is largely unaffected by simply rounding the embedding layer weights to their closest 4-bit values, so we did not apply QAT for the embedding quantization. The embeddings were quantized with a group-size of 32.

\subsection{Unembedding layer pruning}
\LGint{} is trained to generate 128k output tokens, out of which only 20 tokens are used: 
\begin{itemize}
    \item \textbf{Moderation tokens}: 'safe', 'unsafe'
    \item \textbf{Unsafe categories}: '1', '2', '3', '4', '5', '6', '7', '8', '9', '10', '11', '12', '13', '14'
    \item \textbf{Extras}: '\textbackslash n', 'S', ',', '<|eot\_id|>'
\end{itemize}
By keeping only the model connections corresponding to those 20 tokens in the unembedding linear layer,  we can reduce the output layer size significantly without impacting the model. Using this unembedding layer pruning, we reduced the output layer size from 262.6M parameters $(2048\times128k)$ to 40.96k parameters $(2048\times20)$, giving us a total savings of 131.3MB, assuming 4-bit weight quantization, and a final model size of 0.4GB. Although the pruned output layer only generates 20 tokens, they are expanded back to produce the original 128k outputs in the model such that the model interface does not change from the developer’s point of view.

\subsection{Distillation}
We employ \LGeight \citep{dubey2024llama3herdmodels} as a teacher to distill \LGint{} \citep{hinton2015distilling}:
\begin{align}
    \min_{\theta} E_{\mathcal{D}}\left[\mathcal{L}_\text{cross-entropy}\left(y_{s}, y_{t} \right) \right]
\end{align}
where $\theta$ represents the \LGint{} weights, and $y_{s}$ and $y_{t}$ are the student (\LGint{}) and teacher (\LGeight) logits, respectively, for a particular token index. We observe that distillation significantly enhances the model's ability to learn safe and unsafe patterns, as well as the posterior distribution over unsafe categories. We observed a 1.3 percent (absolute) improvement in F1-score on internal test data when using distillation to train \LGint{} compared to standard finetuning.

\section{Results}

\begin{table}\caption{Quantitative comparison between models}\label{table:results}
\resizebox{\columnwidth}{!}{%
\begin{tabular}{|l|c|c|c|c|c|c|c|c|c|c|c|}\hline
\multicolumn{1}{|c|}{{Model}} & \multicolumn{11}{|c|}{F1/FPR}                                                                                                                                                                                                                                                                                                                                                                                                                        \\ \hline
\multicolumn{1}{|c|}{}                       & \multicolumn{1}{|l|}{English} & \multicolumn{1}{|l|}{French} & \multicolumn{1}{|l|}{German} & \multicolumn{1}{|l|}{Italian} & \multicolumn{1}{|l|}{Spanish} & \multicolumn{1}{|l|}{Portugese} & \multicolumn{1}{|l|}{Hindi} & \multicolumn{1}{|l|}{Vietnamese} & \multicolumn{1}{|l|}{Indonesian} & \multicolumn{1}{|l|}{\# param (B)} & \multicolumn{1}{l|}{Model size (GB)} \\ \hline
Llama Guard 3-8B                           & 0.939/0.040                 & 0.943/0.036                & 0.877/0.032                & 0.873/0.038                 & 0.875/0.023                 & 0.860/0.060                   & 0.871/0.050               & 0.890/0.034                    & 0.915/0.048                    & 8                                                                               & 14.9                                                                           \\
Llama Guard 3-1B                           & 0.899/0.090                 & 0.939/0.012                & 0.851/0.06                 & 0.897/0.111                 & 0.84/0.084                  & 0.798/0.108                   & 0.815/0.088               & 0.819/0.099                    & 0.875/0.083                    & 1.5                                                                             & 2.8                                                                            \\
Llama Guard 3-1B -INT4                     & 0.904/0.084                 & 0.873/0.072                & 0.847/0.135                & 0.897/0.111                 & 0.855/0.084                 & 0.844/0.112                   & 0.782/0.104               & 0.825/0.130                    & 0.833/0.121                    & 1.1                                                                             & 0.4                                                                            \\
GPT4                                       & 0.805/0.152                 & 0.795/0.157                & 0.691/0.123                & 0.753/0.20                  & 0.711/0.169                 & 0.738/0.207                   & 0.709/0.206               & 0.741/0.148                    & 0.787/0.169                    & ?                                                                               & ?  \\ \hline                                                                           
\end{tabular}}
\end{table}

As described in \ref{training and evaluation}, we evaluate models using an internal dataset based on \citet{mlcommons} hazard taxonomy. Table \ref{table:results} provides a quantitative comparison between our \LGint{}, \LG{}, the much larger Llama Guard 3-8B, and GPT4. The metrics used in the table are F1 and FPR, which stand for F1 score and False Positive Rate. F1 is a harmonic mean of precision and recall, which means it balances the importance of both metrics and FPR measures the proportion of false positives among all negative instances. For all models other than \LGint{}, we assume the bf16 data format when calculating the model size. \LGint{} achieves better F1/FPR scores for English relative to \LG{} despite having $7\times$ smaller model size. In addition, \LGint{} achieves on par or higher F1 score than \LG{} on 5 out of 8 non-English languages. Compared to GPT4, \LGint{} achieves considerably better F1/FPR for English as well as 7/8 non-English languages. Note that GPT4 was tested in a zero-shot manner (using an internal dataset without any additional fine-tuning or adaptation).

To validate that \LGint{} can be run on a commodity mobile device, we deployed the model to a Moto-Razor phone and observed >= 30 token/s and <=2.5s time-to-first-token.

\section{Limitations}
There are some limitations associated with \LGint{}. First, \LGint{} itself is an LLM fine-tuned from Llama 3.2. Thus, its performance (e.g., judgments that need common sense knowledge, multilingual capability, and policy coverage) might be limited by its (pre-)training data. Llama Guard performance varies across model size and languages. When possible, developers should consider Llama Guard 3-8B which may provide better safety classification performance but comes at a higher deployment cost. Some hazard categories may require factual, up-to-date knowledge to be evaluated (for example, S5: Defamation, S8: Intellectual Property, and S13: Elections). We believe more complex systems should be deployed to accurately moderate these categories for use cases highly sensitive to these types of hazards, but \LGint{} provides a good baseline for generic use cases. Lastly, as an LLM, \LGint{} may be susceptible to adversarial attacks or prompt injection attacks that could bypass or alter its intended use. Please \href{https://github.com/meta-llama/PurpleLlama}{report} vulnerabilities and we will look to incorporate improvements in future versions of Llama Guard.

\section{Summary}
In conclusion, the pruned and quantized Llama Guard 3-1B-INT4 model is a significant improvement over its predecessors in terms of both safety and efficiency. Additionally, the model's small size and low latency make it well-suited for deployment on mobile devices.

\clearpage
\newpage
\bibliographystyle{assets/plainnat}
\bibliography{paper}

\begin{thebibliography}{22}
\providecommand{\natexlab}[1]{#1}
\providecommand{\url}[1]{\texttt{#1}}
\expandafter\ifx\csname urlstyle\endcsname\relax
  \providecommand{\doi}[1]{doi: #1}\else
  \providecommand{\doi}{doi: \begingroup \urlstyle{rm}\Url}\fi

\bibitem[lla(2024)]{llama32}
Llama 3.2: Revolutionizing edge ai and vision with open, customizable models, 2024.
\newblock \url{https://ai.meta.com/blog/llama-3-2-connect-2024-vision-edge-mobile-devices/}.

\bibitem[Bai et~al.(2022)Bai, Jones, Ndousse, Askell, Chen, DasSarma, Drain, Fort, Ganguli, Henighan, Joseph, Kadavath, Kernion, Conerly, El-Showk, Elhage, Hatfield-Dodds, Hernandez, Hume, Johnston, Kravec, Lovitt, Nanda, Olsson, Amodei, Brown, Clark, McCandlish, Olah, Mann, and Kaplan]{bai2022traininghelpfulharmlessassistant}
Yuntao Bai, Andy Jones, Kamal Ndousse, Amanda Askell, Anna Chen, Nova DasSarma, Dawn Drain, Stanislav Fort, Deep Ganguli, Tom Henighan, Nicholas Joseph, Saurav Kadavath, Jackson Kernion, Tom Conerly, Sheer El-Showk, Nelson Elhage, Zac Hatfield-Dodds, Danny Hernandez, Tristan Hume, Scott Johnston, Shauna Kravec, Liane Lovitt, Neel Nanda, Catherine Olsson, Dario Amodei, Tom Brown, Jack Clark, Sam McCandlish, Chris Olah, Ben Mann, and Jared Kaplan.
\newblock Training a helpful and harmless assistant with reinforcement learning from human feedback, 2022.
\newblock \url{https://arxiv.org/abs/2204.05862}.

\bibitem[Bengio et~al.(2013)Bengio, L{\'e}onard, and Courville]{bengio2013estimating}
Yoshua Bengio, Nicholas L{\'e}onard, and Aaron Courville.
\newblock Estimating or propagating gradients through stochastic neurons for conditional computation.
\newblock \emph{arXiv preprint arXiv:1308.3432}, 2013.

\bibitem[Elhage et~al.(2023)Elhage, Lasenby, and Olah]{residualstream}
Nelson Elhage, Roberb Lasenby, and Christopher Olah.
\newblock Privileged bases in the transformer residual stream, 2023.
\newblock \url{https://transformer-circuits.pub/2023/privileged-basis/index.html}.

\bibitem[{Executorch Team}(2024{\natexlab{a}})]{executorch_android}
{Executorch Team}.
\newblock Executorch llama android demo app.
\newblock \url{https://github.com/pytorch/executorch/tree/main/examples/demo-apps/android/LlamaDemo}, 2024{\natexlab{a}}.

\bibitem[{Executorch Team}(2024{\natexlab{b}})]{executorch_ios}
{Executorch Team}.
\newblock Executorch llama ios demo app.
\newblock \url{https://github.com/pytorch/executorch/tree/main/examples/demo-apps/apple_ios/LLaMA}, 2024{\natexlab{b}}.

\bibitem[Hinton(2015)]{hinton2015distilling}
Geoffrey Hinton.
\newblock Distilling the knowledge in a neural network.
\newblock \emph{arXiv preprint arXiv:1503.02531}, 2015.

\bibitem[Inan et~al.(2023)Inan, Upasani, Chi, Rungta, Iyer, Mao, Tontchev, Hu, Fuller, Testuggine, and Khabsa]{inan2023llamaguardllmbasedinputoutput}
Hakan Inan, Kartikeya Upasani, Jianfeng Chi, Rashi Rungta, Krithika Iyer, Yuning Mao, Michael Tontchev, Qing Hu, Brian Fuller, Davide Testuggine, and Madian Khabsa.
\newblock Llama guard: Llm-based input-output safeguard for human-ai conversations, 2023.
\newblock \url{https://arxiv.org/abs/2312.06674}.

\bibitem[Krishnamoorthi(2018)]{krishnamoorthi2018quantizing}
Raghuraman Krishnamoorthi.
\newblock Quantizing deep convolutional networks for efficient inference: A whitepaper.
\newblock \emph{arXiv preprint arXiv:1806.08342}, 2018.

\bibitem[Liu et~al.(2023)Liu, Oguz, Zhao, Chang, Stock, Mehdad, Shi, Krishnamoorthi, and Chandra]{liu2023llm}
Zechun Liu, Barlas Oguz, Changsheng Zhao, Ernie Chang, Pierre Stock, Yashar Mehdad, Yangyang Shi, Raghuraman Krishnamoorthi, and Vikas Chandra.
\newblock Llm-qat: Data-free quantization aware training for large language models.
\newblock \emph{arXiv preprint arXiv:2305.17888}, 2023.

\bibitem[{Llama Team}(2024{\natexlab{a}})]{dubey2024llama3herdmodels}
{Llama Team}.
\newblock The llama 3 herd of models, 2024{\natexlab{a}}.
\newblock \url{https://arxiv.org/abs/2407.21783}.

\bibitem[{Llama Team}(2024{\natexlab{b}})]{metallamaguard2}
{Llama Team}.
\newblock Meta llama guard 2.
\newblock \url{https://github.com/meta-llama/PurpleLlama/blob/main/Llama-Guard2/MODEL_CARD.md}, 2024{\natexlab{b}}.

\bibitem[Llama~Team(2024)]{metallamaguard3}
AI~@~Meta Llama~Team.
\newblock The llama 3 family of models.
\newblock \url{https://github.com/meta-llama/PurpleLlama/blob/main/Llama-Guard3/1B/MODEL_CARD.md}, 2024.

\bibitem[Men et~al.(2024)Men, Xu, Zhang, Wang, Lin, Lu, Han, and Chen]{men2024shortgptlayerslargelanguage}
Xin Men, Mingyu Xu, Qingyu Zhang, Bingning Wang, Hongyu Lin, Yaojie Lu, Xianpei Han, and Weipeng Chen.
\newblock Shortgpt: Layers in large language models are more redundant than you expect, 2024.
\newblock \url{https://arxiv.org/abs/2403.03853}.

\bibitem[{MLCommons}(2024)]{mlcommons}
{MLCommons}.
\newblock Announcing mlcommons ai safety v0.5 proof of concept.
\newblock \url{https://mlcommons.org/2024/04/mlc-aisafety-v0-5-poc/}, 2024.

\bibitem[Muralidharan et~al.(2024)Muralidharan, Sreenivas, Joshi, Chochowski, Patwary, Shoeybi, Catanzaro, Kautz, and Molchanov]{muralidharan2024compactlanguagemodelspruning}
Saurav Muralidharan, Sharath~Turuvekere Sreenivas, Raviraj Joshi, Marcin Chochowski, Mostofa Patwary, Mohammad Shoeybi, Bryan Catanzaro, Jan Kautz, and Pavlo Molchanov.
\newblock Compact language models via pruning and knowledge distillation, 2024.
\newblock \url{https://arxiv.org/abs/2407.14679}.

\bibitem[Nagel et~al.(2021)Nagel, Fournarakis, Amjad, Bondarenko, Van~Baalen, and Blankevoort]{nagel2021white}
Markus Nagel, Marios Fournarakis, Rana~Ali Amjad, Yelysei Bondarenko, Mart Van~Baalen, and Tijmen Blankevoort.
\newblock A white paper on neural network quantization.
\newblock \emph{arXiv preprint arXiv:2106.08295}, 2021.

\bibitem[{Pytorch Team}(2024{\natexlab{a}})]{executorch}
{Pytorch Team}.
\newblock Executorch runtime overview.
\newblock \url{https://pytorch.org/executorch/stable/runtime-overview.html}, 2024{\natexlab{a}}.

\bibitem[{Pytorch Team}(2024{\natexlab{b}})]{xnnpack}
{Pytorch Team}.
\newblock Executorch xnnpack delegate.
\newblock \url{https://pytorch.org/executorch/stable/native-delegates-executorch-xnnpack-delegate.html}, 2024{\natexlab{b}}.

\bibitem[Team et~al.(2024)Team, Mesnard, Hardin, Dadashi, Bhupatiraju, Pathak, Sifre, Rivi{\`e}re, Kale, Love, et~al.]{team2024gemma}
Gemma Team, Thomas Mesnard, Cassidy Hardin, Robert Dadashi, Surya Bhupatiraju, Shreya Pathak, Laurent Sifre, Morgane Rivi{\`e}re, Mihir~Sanjay Kale, Juliette Love, et~al.
\newblock Gemma: Open models based on gemini research and technology.
\newblock \emph{arXiv preprint arXiv:2403.08295}, 2024.

\bibitem[torchao maintainers and contributors(2024)]{torchao}
torchao maintainers and contributors.
\newblock torchao: Pytorch native quantization and sparsity for training and inference, October 2024.
\newblock \url{https//github.com/pytorch/torchao}.

\bibitem[torchtune maintainers and contributors(2024)]{torchtune}
torchtune maintainers and contributors.
\newblock torchtune: Pytorch's finetuning library, April 2024.
\newblock \url{https//github.com/pytorch/torchtune}.

\end{thebibliography}

\newpage
\beginappendix

\section{Acknowledgements}
This work was made possible by a large group of contributors. We extend our gratitude to the following
people: Jacob Szwejbka, Jack Khuu, Chester Hu, Riandy Riandy, Chirag Modi, Beto de Paola, Jana Vranes, Botao Chen, Kartikeya Upasani, Ujjwal Karn, Madian Khabsa, Varun Vontimitta, Vincent Gonguet, Joe Spisak. 

\end{document}